\input ./style/arxiv-general.cfg
\documentclass[aoas,MSNbibl,nameyear,dvips]{arximspdf}
\makeatletter
   \@ifpackageloaded{graphicx}{}{\usepackage{graphicx}}
\makeatother

\doi{10.1214/15-AOAS867}
\volume{9}
\issue{4}
\pubyear{2015}
\firstpage{1932}
\lastpage{1949}
\docsubty{FLA}

\makeatletter

\newcommand{\Sigmav}{\bolds{\Sigma}}
\newcommand{\av}{\mathbf{a}}
\makeatother

\begin{document}
\begin{frontmatter}

\title{Lymphangiogenesis and carcinoma in the uterine cervix: Joint and
hierarchical models for random cluster sizes and continuous outcomes}
\runtitle{Models for random cluster sizes and continuous outcomes}

\begin{aug}
\author[A]{\fnms{T.~R.}~\snm{Fanshawe}\corref{}\thanksref{m1}\ead[label=e1]{thomas.fanshawe@phc.ox.ac.uk}},
\author[B]{\fnms{C.~M.} \snm{Chapman}\thanksref{m2}\ead[label=e2]{Mark.Chapman@mbht.nhs.uk}}
\and
\author[B]{\fnms{T.}~\snm{Crick}\thanksref{m2}\ead[label=e3]{tony.crick@mbht.nhs.uk}}
\runauthor{T.~R. Fanshawe, C.~M. Chapman and T. Crick}
\affiliation{University of Oxford\thanksmark{m1} and Royal Lancaster
Infirmary\thanksmark{m2}}
\address[A]{T.~R. Fanshawe\\
Nuffield Department of Primary Care\\
\quad Health Sciences\\
University of Oxford\\
Radcliffe Observatory Quarter\\
Woodstock Road, Oxford OX2 6GG\\
United Kingdom\\
\printead{e1}}
\address[B]{C.~M. Chapman\\
T. Crick\\
Department of Histopathology\\
Royal Lancaster Infirmary\\
Ashton Road, Lancaster\\
Lancashire LA1 4RP\\
United Kingdom\\
\printead{e2}\\
\phantom{E-mail:\ }\printead*{e3}}
\end{aug}

%
\received{\smonth{6} \syear{2013}}
%
\revised{\smonth{7} \syear{2014}}

%
\begin{abstract}
Although the lymphatic system is clearly linked to the metastasis of
most human carcinomas, the mechanisms by which lymphangiogenesis occurs
in response to the presence of carcinoma remain unclear. Hierarchical
models are presented to investigate the properties of lymphatic vessel
production in 2997 fields taken from 20 individuals with invasive
carcinoma, 21 individuals with cervical intraepithelial neoplasia and
21 controls. Such data demonstrate a high degree of correlation within
tumour samples from the same individual. Joint hierarchical models
utilising shared random effects are discussed and fitted in a Bayesian
framework to allow for the correlation between two key outcome
measures: a random cluster size (the number of lymphatic vessels in a
tissue sample) and a continuous outcome (vessel size). Results show
that invasive carcinoma samples are associated with increased
production of smaller and more irregularly-shaped lymphatic vessels and
suggest a mechanistic link between carcinoma of the cervix and
lymphangiogenesis.
\end{abstract}

%
\end{frontmatter}

\section{Introduction}\label{sec1}

Observational and randomized studies often provide data with a
multilevel, or hierarchical, structure, in which repeated data values
are available in ``clusters'' at one level of the hierarchy. Each subunit
contributes for data analysis a certain number of observations, which
might vary across clusters and which might therefore be regarded as a
random variable---a ``cluster-specific sample size'' or, simply ``cluster
size''. In recent years, consideration has been given to the issue of
so-called ``informative cluster size'', in which the number of
observations within a cluster is associated with a study outcome.

This issue is potentially important in many application areas.
Prominent amongst them is the field of developmental toxicity, where,
for example, a correlation has been shown between animal litter size
and animal-specific outcomes such as malformation or birthweight,
giving rise to a series of papers [\citet
{fitzmaurice,tenhave,regan,dunson,gueorguieva,ma}]. Another example
arises in periodontics, where there may be an association between tooth
loss and tooth quality. This is explored by \citet{williamson} and
\citet
{neuhaus}; the latter paper also considers a parallel between the\vadjust{\goodbreak}
generic problem of informative cluster size and informative drop-out in
longitudinal studies. Further examples appear in educational research
[class size and examination performance, \citet{goldstein}] and human
perinatal epidemiology [multiple births and various outcomes, \citet
{hibbs}]. The general methodological approach has also been extended to
survival analysis outcomes [\citet{cong}].

This paper introduces the issue of informative cluster size in the
analysis of histological data taken from uterine cervical carcinoma
samples. Carcinoma of the uterine cervix is the second most common
malignant neoplasm amongst females globally, and in 2008 almost half a
million individuals were diagnosed with this condition [\citet{ferlay}].
The preferred route of metastasis (``spread'') for carcinomas is via the
lymphatic system [\citet{friedl}]. Studies into the role of the
lymphatic system in the progression of cervical carcinoma have
demonstrated that the density of lymphatic vessels (LVD) is a good
indicator of lymph node metastasis, higher tumour grades and lymphatic
invasion [\citet{gombos,gao,longatto,zhangyuzhang}].

Moreover, these studies suggest that cervical tumours have the ability
to induce lymphangiogenesis, the formation of new lymphatic vessels,
but provide little information regarding the distribution of LVD in
normal cervix and premalignant conditions, and often fail to detail the
anatomical cervical location in which LVD is measured. As 90\% of all
cervical lesions occur in the region known as the transformation zone,
any difference in the LVD of this anatomical region compared with the
other regions of the cervix (the ectocervix and endocervix) is of
particular importance. Together this information may help determine at
what stage in the progression of the disease lymphangiogenesis takes place.

Previous studies have made observations describing the morphological
appearance of lymphatic vessels in cervical tissue. For example, \citet
{gombos}, \citet{gao} and \citet{zhangyuzhang} observed that the
lymphatics in normal cervical tissue appear open with regular shape,
whilst those in the peritumoral regions of carcinoma tissue appear
large and dilate. To build\vadjust{\goodbreak} on this observational data, the present
study aims to utilise quantitative data obtained from image analysis to
describe the number, size and shape of lymphatic vessels in the uterine
cervix via measurements of LVD, vessel area and circularity.

Lymphangiogenesis is thought to occur via the sprouting of endothelial
cells from existing lymphatic vessels [Alitalo, Tammela and Petrova
(\citeyear{al05})]. If this is the case, a subset of smaller lymphatic vessels may
be visible in tissue from carcinoma specimens. The structural
arrangement of these newly formed vessels in 3D space will influence
how functional they are as compared to those found in the normal
cervix. This study addresses these issues.

In the context of the more general issue of informative cluster size
described above, the methodological challenge arises when further
outcome variables---such as the size of each vessel---are associated
with the number of vessels at a particular level of observation. For
example, in an inverse relationship such as the one described in this
paper, large clusters contain vessels that tend to be smaller in
magnitude than those in small clusters, yet by their nature the large
clusters provide more measurements for analysis. Ignoring the effects
of clustering and cluster size may then provide incorrect inferences
about the outcome variable, and assessing the extent to which this is
true is also one of the objectives of this paper.

Scientific interest in this study therefore lies not only in the number
of lymphatic vessels observed, but also in quantitative measures of
their appearance. This objective naturally suggests a joint modelling
approach. A key requirement is that the model must be flexible enough
to allow not only for the correlation among these outcome variables,
but also for the fact that they may differ in statistical distribution
or data type (e.g., count as opposed to continuous) and at the level of
the hierarchy at which they are measured.

In this paper we illustrate the general modelling approach by
concentrating on the relationship between LVD and vessel size, adapting
for our application methodology that has been successfully applied in
fields such as toxicology [\citet{regan}]. Our modelling approach relies
on the specification of random effects that are common to both outcome
variables in the model. These random effects provide a mechanism by
which the correlation between the outcomes can be modelled explicitly.

The paper is structured as follows. In Section~\ref{sec2} we describe in detail
the data set that motivates this work, and give an exploratory
analysis. In Section~\ref{hierarchical} we fit univariate hierarchical models to each of
the outcome variables of interest. In Section~\ref{joint} we introduce the joint
modelling problem and present a bivariate model for lymphatic vessel
density and lymphatic vessel area, and Section~\ref{sec5} provides a concluding
discussion.\vadjust{\eject}
\section{Data}\label{sec2}

\subsection{Study design}\label{sec2.1}

The data were collected as part of a study carried out at the Royal
Lancaster Infirmary, Lancaster, UK. Tissue biopsies (or ``specimens'')
were taken from 62 individuals. Each specimen was processed into
paraffin blocks, which were sectioned at 4~$\mu$m, stained and viewed
under a microscope, as described by \citet{chapman}. Within each
specimen, areas of interest (``fields'') were selected and all lymphatic
vessels observed within these fields were used to obtain the outcomes
of interest, defined below.

Twenty individuals provided invasive squamous cell carcinoma tissue,
while 21 individuals showed premalignant growth classified as cervical
intraepithelial neoplasia (CIN), which was additionally subclassified
as histological grade 1, 2 or 3 (Table~\ref{specimentable}). For these
two groups of cases, fields were taken from the site of abnormal
growth. Additionally, hysterectomy specimens were obtained from 21
controls---defined as individuals with menorrhagia, with no abnormal
cervical tissue. For the controls, specimens were available either from
one or, more commonly, from two distinct functional regions of the
cervix---the ectocervix and the transformation zone.

An average of 8.8 fields were taken per specimen (range 1 to 19 fields
per specimen), and across the whole sample an average of 5.5 lymphatic
vessels per field provided data for analysis (range 1 to 45 vessels per
field). The reasons for apparent differences by group in the number of
fields per specimen shown in Table~\ref{specimentable} are unrelated to
outcome variables.

\begin{table}
\caption{Number of specimens and fields per specimen}
\label{specimentable}
\begin{tabular*}{\textwidth}{@{\extracolsep{\fill}}lcc@{}}
\hline
 &  & \textbf{Average number of fields}\\
\textbf{Tissue type}& \textbf{Number of specimens} & \textbf{per specimen (range)}\\
\hline
Control cervix & 21 & 12.7 (5 to 19)\\
\quad Ectocervix & 20 & \phantom{0}9.2 (5 to 10)\\
\quad Transformation zone & 16 & 5.2 (2 to 9)\\[3pt]
CIN & 21 & 5.6 (2 to 8)\\
\quad CIN1 & 10 & 5.4 (2 to 7)\\
\quad CIN2 & \phantom{0}9 & 6.1 (2 to 8)\\
\quad CIN3 & \phantom{0}2 & 4.5 (4 to 5)\\[3pt]
Invasive carcinoma & 20 & \phantom{0}8.0 (1 to 10)\\
\hline
\end{tabular*}
\end{table}

Two field-level outcomes and two vessel-level outcomes were of primary interest:
\begin{itemize}
\item Lymphatic vascular density (LVD)---the density of lymphatic
vessels visible in a field.
\item Percentage lymphatic area (\%LA)---the percentage of the total
area of a field that is occupied by lymphatic vessels.
\item Vessel area---the area contained within the lumen of a lymphatic
vessel, measured in $\mu\mathrm{m}^2$.
\item Circularity---a measure of the circularity of a lymphatic vessel,
lying between 1 (perfectly circular) and 0 (lying in parallel lines
across the surface of the field).
\end{itemize}

Note that as all fields were of the same area, LVD is almost equivalent
to the number of lymphatic vessels in a field: a discrepancy would
arise only if a convoluted vessel were visible at two or more distinct
points on the same field, a scenario that is impossible to detect using
the data available and which we consequently ignore. The \%LA of a
field can be viewed as a combined summary measure of the LVD and the
average vessel area of the field, while remaining an important outcome
variable in its own right.

A sample size calculation was carried out based on analysis of variance
to test for a difference in mean LVD (calculated across all fields,
averaging to remove the hierarchical structure) between the three main
study groups.
\citet{longatto} and \citet{gao} provide information on LVD in previously
conducted studies, although both studies purposively oversampled
regions of high LVD. \citet{longatto} report mean LVD values of 2.6 in
the control group (ignoring the distinction between ectocervix and
transformation zone), 5.0 for patients with squamous intraepithelial
lesions and 17.1 for patients with invasive carcinoma. Because of
concerns about the nature of the sampling scheme in these studies, the
present study instead assumed more conservative mean LVD values in the
three groups of 2.6, 5.0 and 10.0, respectively, which correspond to a
between-group standard deviation of 3.8. Based on further results
provided by \citet{longatto}, the common within-group standard deviation
was assumed to be 7.5, which yields a ``difference parameter'' [\citet
{day}] of $3.8/7.5 \approx0.5$. Under these assumptions, a sample size
of 25 patients per group has approximately 90\% power to detect an
overall difference in mean LVD between groups at the 5\% level of
significance. Incorporating repeated measurements from different fields
into the analysis is likely to increase the power substantially.

\begin{table}
\caption{Mean (standard deviation) of outcome
measures by tissue type}
\label{summarytable}
\begin{tabular*}{\textwidth}{@{\extracolsep{\fill}}lcccc@{}}
\hline
& \textbf{LVD} & \textbf{\%LA} & \textbf{Vessel area} & \textbf{Circularity}\\
\hline
Control cervix & 3.38 (2.17) & 4.05 (3.66) & 1633 (2555) & 0.57 (0.21)\\
\quad Ectocervix & 2.34 (0.90) & 3.51 (3.31) & 2062 (2764) & 0.54
(0.21)\\
\quad Transformation zone & 5.67 (2.39) & 5.22 (4.11) & 1248 (2285) &
0.60 (0.21)\\[3pt]
CIN & 5.53 (1.75) & 3.76 (3.09) & \phantom{0}912 (1489) & 0.61 (0.19)\\
\quad CIN1 & 5.33 (1.81) & 3.94 (3.72) & 1010 (1851) & 0.59 (0.20)\\
\quad CIN2 & 5.91 (1.67) & 3.90 (2.54) & \phantom{0}869 (1166) & 0.63 (0.19)\\
\quad CIN3 & 4.44 (1.24) & 1.90 (0.84) & \phantom{0}543 (540) & 0.64 (0.17)\\[3pt]
Invasive carcinoma & 9.04 (4.55) &3.47 (2.20) &  \phantom{0}523 (934) & 0.56
(0.22)\\
\hline
\end{tabular*}
\end{table}

\subsection{Exploratory analysis}\label{sec2.2}

\begin{figure}

\includegraphics{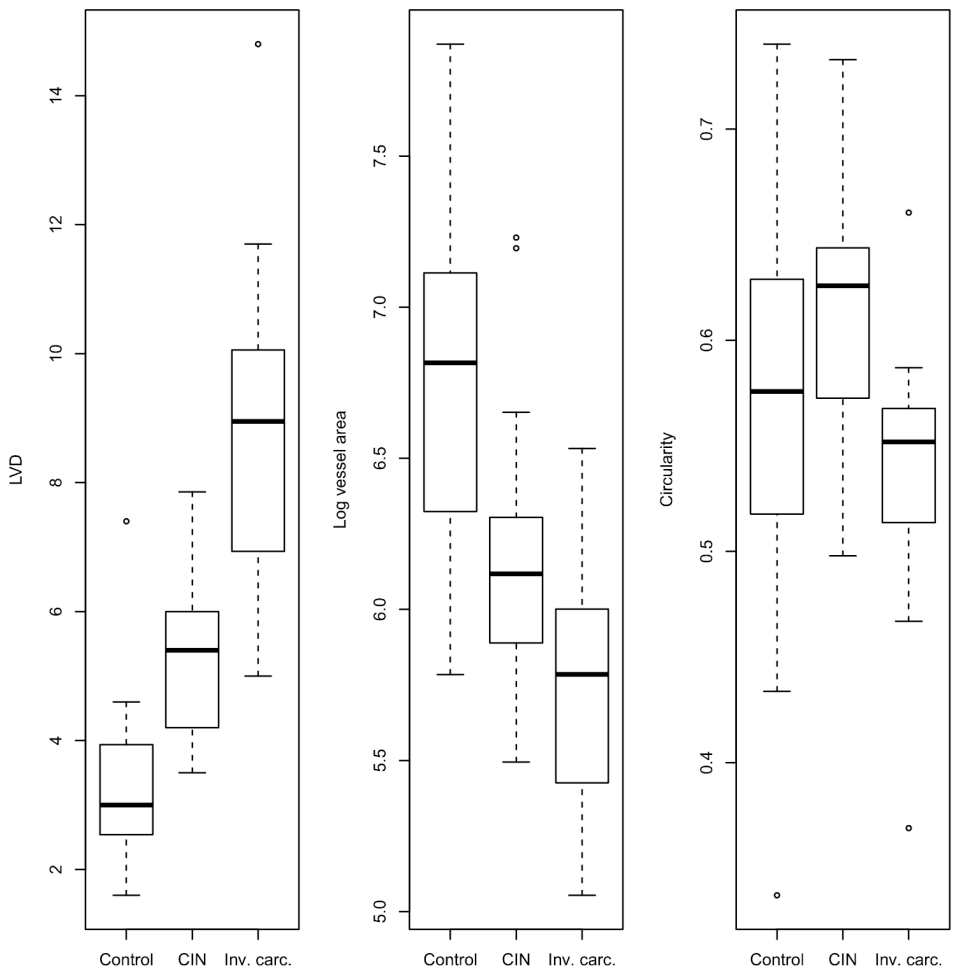}

\caption{Distributions of three outcome variables by tissue type. ``Inv.
carc.'' is invasive carcinoma.}
\label{boxplot}
\end{figure}

Table~\ref{summarytable} shows the means and standard deviations of the
outcome measures by group. The distributions of three of these
variables are illustrated by Figure~\ref{boxplot}, in which the two
control tissue groups are combined, as are the three CIN groups. Vessel
area has an extremely positively skewed distribution and is therefore
presented on the logarithmic scale.

\begin{figure}

\includegraphics{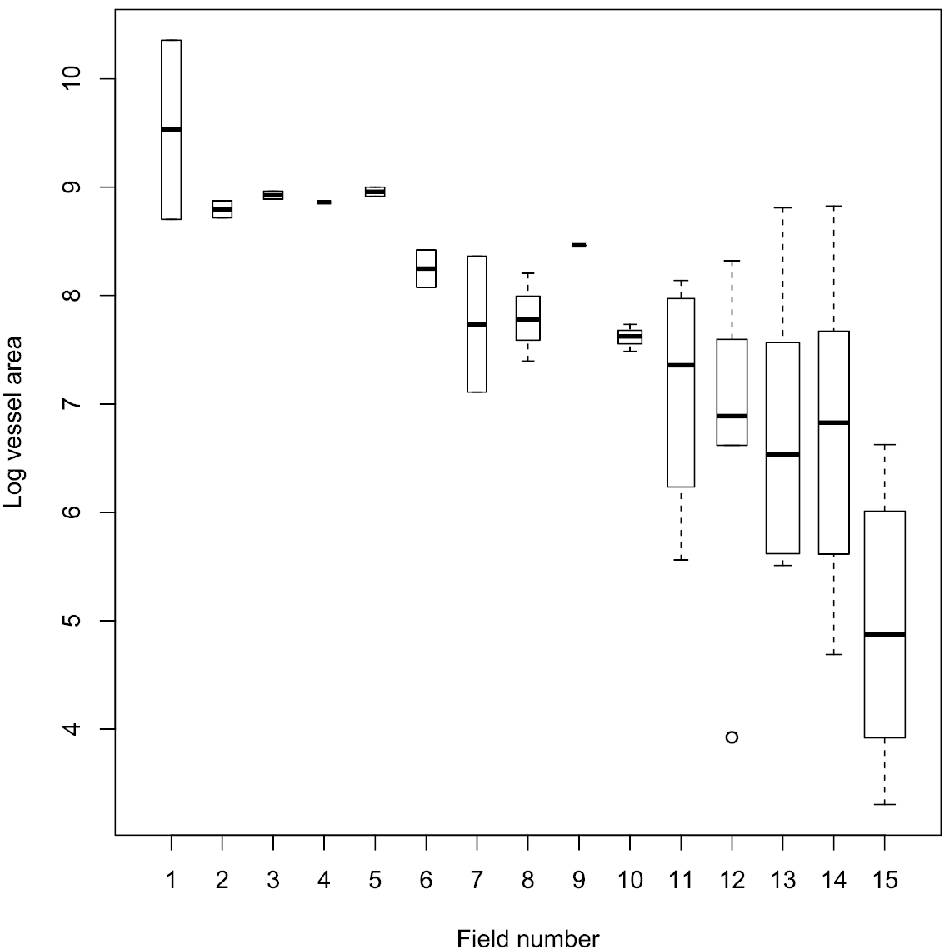}

\caption{Variation in log-vessel area for the fifteen fields taken from
a single specimen (the width of each box is proportional to the square
root of the number of vessels).}
\label{fieldvariation}
\end{figure}

\begin{figure}

\includegraphics{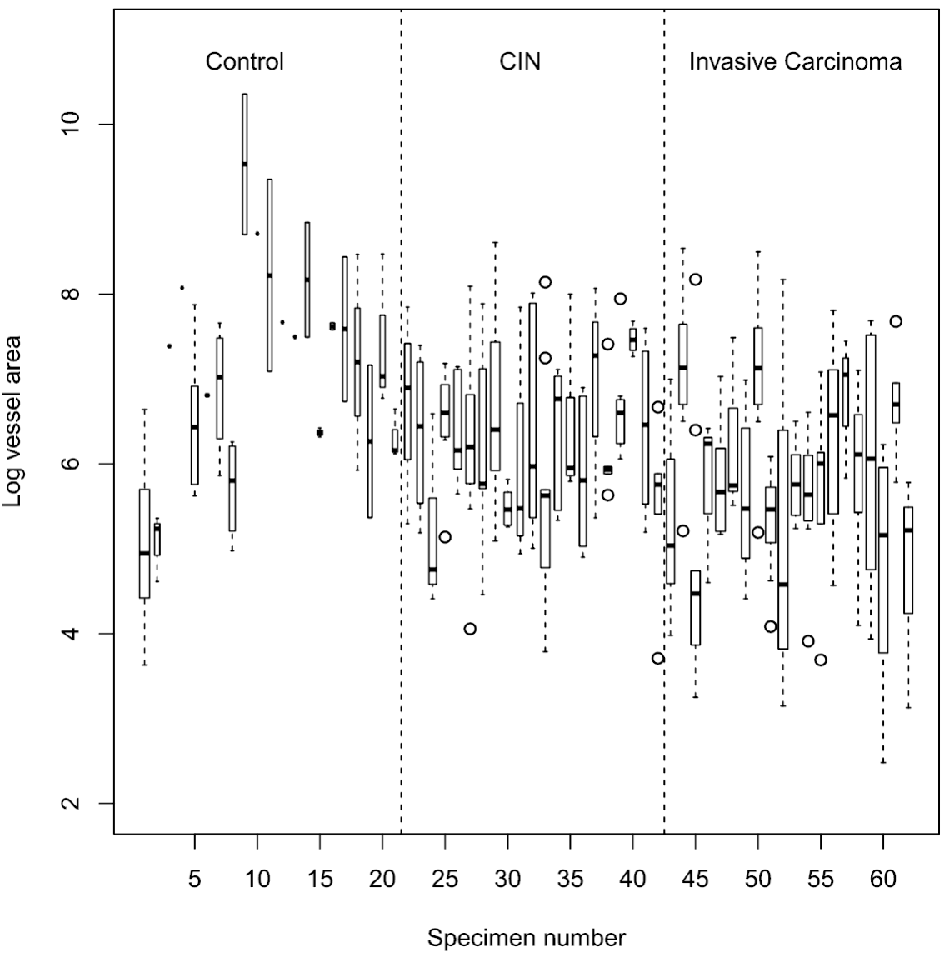}

\caption{Variation in log-vessel area for the first-numbered fields
taken from each specimen (the width of each box is proportional to the
square root of the number of vessels).}
\label{specimenvariation}
\end{figure}

Two key further considerations guide our approach to analysing this
data-set. Firstly, outcomes vary according to the level of the
hierarchy at which they are observed---fields within specimens, and
vessels within fields. Figure~\ref{fieldvariation} provides a typical
example, and shows the log-vessel area of vessels in the 15 fields
taken from the same specimen. In a similar vein, Figure~\ref{specimenvariation} shows the variation in log-vessel area of vessels
in the first-numbered fields taken from each specimen.

\begin{figure}

\includegraphics{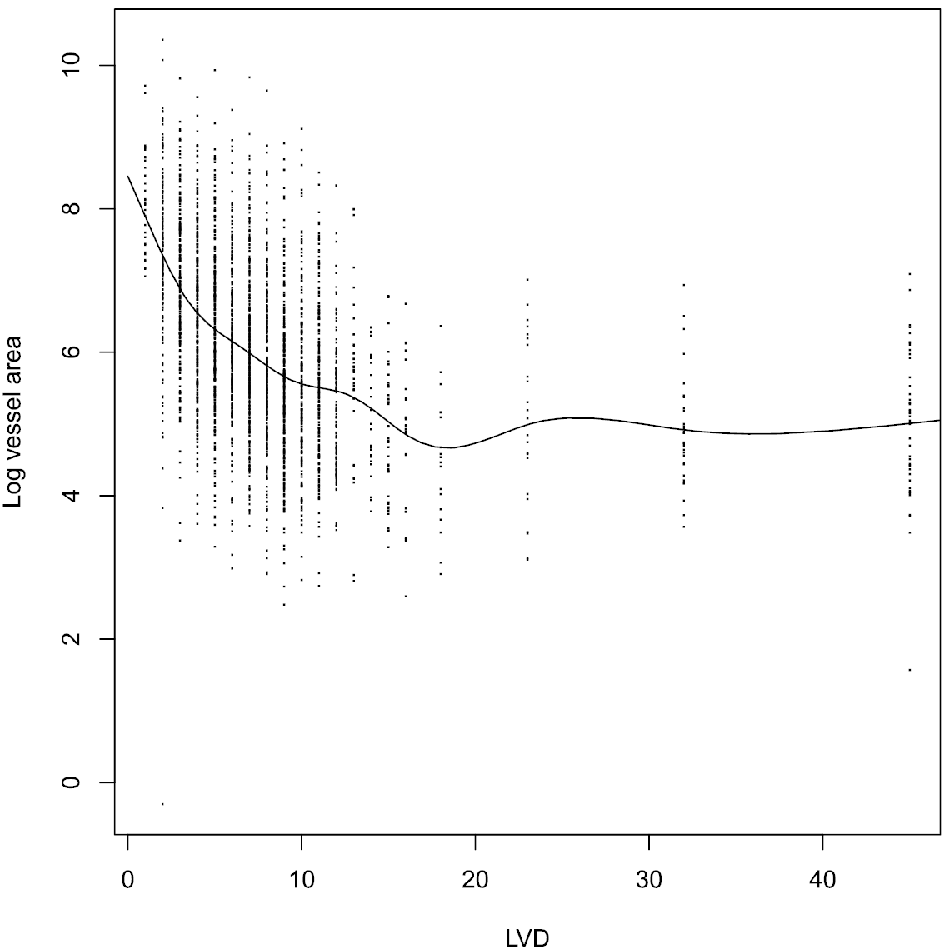}

\caption{Relationship between log-vessel area and LVD for all vessels,
with fitted loess curve.}
\label{VAversusLVD}
\end{figure}

The second major consideration is that the outcome variables themselves
are correlated, notably the important variables LVD and vessel area.
Figure~\ref{VAversusLVD} plots log-vessel area and LVD across all
vessels, together with the fitted curve resulting from a generalised
additive model fit to the data ignoring the hierarchical structure. The
sample correlation between the two is $-0.36$, with stronger correlation
apparent in fields with fewer than fifteen vessels.

\section{Hierarchical models}
\label{hierarchical}

\subsection{Notation}

The data set exhibits a clear three-level hierarchical structure, with
vessels nested within fields, which are themselves nested within
individuals. For specimen $i$, field $j$ and vessel $k$, let $Y_{ij}^{\%
}$ denote the field-level outcome \%LA, $Y_{ijk}^A$ the vessel-level
outcome vessel area, and $Y_{ijk}^C$ the vessel-level outcome
circularity. Additionally, let $N_{ij}$ denote the field-level LVD,
which can be thought of as representing the cluster-specific sample
size. The ranges of the subscripts are $i=1,\ldots,n=62$, $j=1,\ldots,n_i$
and $k=1,\ldots,N_{ij}$.

Note that $n$ (the number of individuals) and the $n_i$ (the number of
fields for individual $i$) are regarded as fixed, and determined by the
study design, whereas $N_{ij}$ is a random variable, an observation
that will be explored further in Section~\ref{joint}. Note also that
the $j$ subscript enumerates fields in both the ectocervix and
transformation zone for the fifteen controls who contribute both of
these tissue types. Let $x_{ij}$ denote the tissue type of specimen $i$
and field $j$, with $x_{ij} \equiv x_i$ for CIN and invasive carcinoma
specimens. In subsequent regression models results are presented
relative to the reference category of control ectocervix. In most cases
the three CIN categories were combined for the purposes of model
fitting because of the small sample size in the CIN3 group and because
differences between the other two categories were small.

The models in this section are conditional on the observed value of
$N_{ij}$, an extremely common approach in the analysis of hierarchical
or multilevel data, albeit one that is often made only implicitly. For
example, in educational research and in many cluster randomised trials
[e.g., \citet{carter}], random cluster sizes are widely discussed, and
considerations such as school size and hospital or ward size in a
health setting might reasonably be expected to demonstrate an
association with outcome measures. In the present study, $N_{ij}$ is
prespecified as a key outcome variable to which a priori
hypotheses attain.

Different models were formulated for each of the four outcome
variables, although each is a variant on the three-level hierarchical
model with independent random effects governing within-level
correlation [\citet{pinheiro}].

\subsection{Models}

For clarity of notation in this section, subscripts are not used to
distinguish between corresponding parameters (such as the intercept
parameter) for different outcomes. Percentage lymphatic area was
modelled untransformed as
%
\begin{equation}
\label{PLAeqn} Y_{ij}^{\%} = \alpha+ \beta_{x_{ij}} +
a_i + Z_{ij},
\end{equation}
where $a_i \sim\mathrm{N}(0,\tau^2)$ independently is a specimen-level
random effect and $Z_{ij} \sim\mathrm{N}(0,\sigma^2)$ is an error
term, where the $Z_{ij}$ are independent of each other and of the
$a_i$, that is, fields are assumed to be conditionally independent,
given the specimen. Unless otherwise stated, similar independence
assumptions are made on random effect and error terms in the other
models described in this section.

The count variable LVD was modelled using a generalised mixed model. An
offset of unity was included to prevent zero counts, as any fields
containing no vessels would not have been included in the data set,
%
\begin{eqnarray}
\label{LVDeqn} N_{ij} & \sim&1 + \operatorname{Poisson}(\mu_{ij}),
\nonumber
\\[-8pt]
\\[-8pt]
\nonumber
\log(\mu_{ij}) & =& \alpha+ \beta_{x_{ij}} + a_i.
\end{eqnarray}

The above formulation can be viewed as part of the general class of
models for overdispersion discussed by \citet{dean}. In order to test
whether this adequately captured the overdispersion in $N_{ij}$,
results were compared to those from a negative binomial model,
following \citet{lindsey}.

The two vessel-level variables were modelled using an extension of the
hierarchy to allow for field-specific random effects. Vessel area was
log-transformed and modelled as
%
\begin{equation}
\label{VAeqn} \log\bigl(Y_{ijk}^A\bigr) = \alpha+
\beta_{x_{ij}} + a_i + b_{ij} + Z_{ijk},
\end{equation}
with $b_{ij} \sim\mathrm{N}(0,\nu^2)$ independently of the $a_i$ and
the $Z_{ijk}$.

The circularity variable $Y_{ijk}^C$ required a somewhat different
approach owing to the substantive hypothesis relating to this variable.
While hypotheses for the other three outcomes all referred to mean
differences between tissue types, the question to be answered using the
circularity data is whether there is a differential in the structure of
lymph cells drawn from specimens belonging to different groups. In
particular, it was hypothesised that lymphatic vessels in control
tissue would retain a more regular structure than those in case tissue;
in this case, control lymphatic vessels might tend to have the
appearance of lying parallel to one another. This question is not
readily answered using mean circularity, which would instead measure
the extent to which lymphatic vessels tend to lie parallel to the plane
at which the biopsy is taken, cutting the three-dimensional tissue at
an arbitrary angle [\citet{wicksell}].

Instead, it is required to test whether the within-specimen,
within-field variation in lymphatic vessel circularity is greater in
cases than in controls, which would indicate a less regularly aligned
lymphatic vessel network. Thus, using a logit transform to transform
the domain of $Y_{ijk}^C$ from $(0,1)$ to $(-\infty,\infty)$, the model
is of the form
%
\begin{equation}
\label{Ceqn} \operatorname{logit}\bigl(Y_{ijk}^C\bigr) = \alpha+
\beta_{x_{ij}} + a_i + b_{ij} + Z_{ijk},
\end{equation}
where random effects are specified as above, except that $b_{ij} \sim
\mathrm{N}(0,\delta_{x_{ij}} \nu^2)$, with $\delta_4$
(corresponding to
invasive carcinoma) set to unity for identifiability, and we require to
test the hypothesis that all of the $\delta$-parameters are equal
against a general alternative.

Models were fitted using functions in the \texttt{nlme} and \texttt
{MASS} packages in \textsf{R} [\citet{nlmecite,MASScite,Rcite}].

\subsection{Results}

The results of fitting models (\ref{PLAeqn})--(\ref{Ceqn}) are
summarised in Table~\ref{resultstable}. For all outcomes, there was
statistically significant clustering within specimens and also within
fields (nonzero random effect variance parameters). Intra-cluster
correlation coefficients can be estimated from the variance parameter
estimates in Table~\ref{resultstable}; for example, the within-specimen
clustering effect for \%LA is estimated as $1.20/(1.20+8.63)=0.12$. For
the two field-level outcomes, clustering effects are higher at the
field-specific level of the hierarchy than within specimens. Standard
errors of most fixed effect estimates are increased by a factor of
around two compared to corresponding models that make no allowance for
the hierarchical data structure, and point estimates remain similar. No
improvement on the Poisson random effects model was seen using the
negative binomial distribution (in terms of change in the
log-likelihood), and so the Poisson formulation was retained.

Compared to control ectocervix tissue, there is clear evidence that all
other tissue types have greater LVD and smaller vessel area. The
difference is especially marked for invasive carcinoma tissue, for
which there is an estimated 3.7-fold increase in LVD and 3.8-fold
reduction in average vessel area compared to control ectocervix. As LVD
and vessel area show opposing trends, as a result of the correlation
between them, there are no differences between normal ectocervix, CIN
and invasive carcinoma tissue for \%LA, although there is evidence that
\%LA is higher in normal transformation zone tissue.

There is evidence that vessels tend to be more circular in control
transformation zone and CIN than in the other two tissue types, but
also evidence that the within-field variance of circularity
measurements is higher in invasive carcinoma specimens than in control
ectocervix and CIN (as estimates of the relevant $\delta$-parameters
are significantly less than unity). Moreover, analysis of variance
comparing the fit of this model with the special case in which all
$\delta$-parameters are constrained to be equal suggests significantly
improved fit of the more general model ($p=0.001$). Allowing a separate
$\delta$-parameter for each of the three CIN categories, however, did
not greatly improve the fit ($p=0.07$), so the more parsimonious model
is reported here. It should be noted that the study was not designed to
investigate differences between the three CIN grades. The final model
thus suggests greater variation in circularity amongst fields taken
from invasive carcinoma tissue than for control ectocervix and CIN.

\begin{table}
\tabcolsep=0pt
\caption{Parameter estimates for hierarchical models}
\label{resultstable}
\begin{tabular*}{\textwidth}{@{\extracolsep{\fill}}lcccc@{}}
\hline
& \textbf{LVD} & \textbf{\%LA} & \textbf{Vessel area} & \textbf{Circularity}\\
& $\bolds{\exp(\hat{\beta})}$ & $\bolds{\hat{\beta}}$ & $\bolds{\exp(\hat{\beta})}$ &
$\bolds{\exp(\hat
{\beta})}$\\
\hline
Control cervix & \\
\quad Ectocervix & -- & -- & -- & -- \\
\quad Transformation zone & 2.37 $[2.11,2.67]$ & 1.85 $[1.05,2.65]$  & 0.53
$[0.43,0.64]$ & 1.27 $[1.08,1.50]$\\[3pt]
CIN & 2.31 $[1.97,2.70]$ & \phantom{0.}0.28 $[-0.70,1.26]$ & 0.42 $[0.32,0.56]$ & 1.37
$[1.11,1.68]$\\[3pt]
Invasive carcinoma & 3.71 $[3.19,4.31]$ & $-0.04$ $[-0.99,0.91]$ & 0.26
$[0.20,0.35]$ & 1.01 $[0.82,1.23]$\\[6pt]
$\hat{\tau}^2$ & 0.03 $[0.02,0.06]$ & 1.20 $[0.61,2.38]$ & 0.12 $[0.07,0.20]$
& 0.05 $[0.03,0.10]$\\
$\hat{\nu}^2$ & -- & -- & 0.22 $[0.17,0.28]$ & 0.13 $[0.09,0.17]$\\
$\hat{\sigma}^2$ & -- & 8.63 $[7.61,9.80]$ & 1.02 $[0.97,1.08]$ & 0.95
$[0.88,1.03]$\\
$\hat{\delta}_1$ & -- & -- & -- & 0.85 $[0.78,0.93]$\\
$\hat{\delta}_2$ & -- & -- & -- & 0.98 $[0.90,1.05]$\\
$\hat{\delta}_3$ & -- & -- & -- & 0.91 $[0.85,0.97]$\\
\hline
\end{tabular*}
\end{table}

\section{Joint models}
\label{joint}

\subsection{Introduction}

The key issue remaining to be addressed relates to the assumption made
implicitly in hierarchical modelling such as in Section~\ref{hierarchical} that the cluster-specific sample sizes are fixed. In the
present study it is reasonable to regard the total sample size, $n$,
and the number of fields per individual, $n_i$, to be fixed by design,
but the number of vessels per field, $n_{ij}$, would be more accurately
regarded as the realisation of a random variable $N_{ij}$, which is the
definition of the LVD outcome variable.

Fitting models that condition on the observed value of this random
variable may lead to incorrect inferences about the relationship
between study group and other outcomes of interest. For example, there
may be underlying tissue-specific characteristics that are associated
with both an increase in LVD and a reduction in vessel size; indeed, it
is biologically plausible that this should be the case in invasive
carcinoma tissue. These effects may be masked by an analysis that is
based on conditioning on LVD, and the conditional models may give
inappropriate inferences about the relationship between tissue type and
vessel size.

\subsection{Models}

One possible analysis strategy factorises the joint likelihood of $Y$
and $N$ into conditional and marginal components, that is, $[Y,N|X] =
[Y|N,X] [N|X]$, where the notation ``$[\cdot]$'' means ``distribution of''. This
has the appealing property of allowing a simple marginal analysis of
$N|X$, but raises the question of whether the models for $Y|N,X$
considered in Section~\ref{hierarchical} are satisfactory even as
conditional models, as they do not explicitly model the correlation
between $N$ and $Y$.

As a possible solution, \citet{catalanoryan1992} discuss bivariate
models in which a function of $N$ enters the expression for $[Y|N,X]$
directly as a covariate. This approach is also adopted by \citet
{panageas} in the context of outcome of surgery when the cluster size
is the number of patients treated for a given surgeon.

Geometrical considerations and Figure~\ref{VAversusLVD} suggest that an
appropriate choice in the present study might be $N^{-1}$ for the
following reasons. In 2D, lymphatic vessels are approximately circular
in appearance. The maximum total area of $m^2$ circles of equal radius
packed into a square of area $a^2$ is $\pi a^2/4m^2$, and so in two
dimensions the proportion of the field area that is filled by vessels
might be expected to be proportional to the reciprocal $N$. Model (\ref
{VAeqn}) would become
\[
\log\bigl(Y_{ijk}^A\bigr) = \alpha+ \beta_{x_{ij}} +
\gamma n_{ij}^{-1} + a_i + b_{ij} +
Z_{ijk}.
\]
This approach causes a manifest change in the parameter estimates: the
estimates of $\exp(\beta)$ change to $1.09\ [0.88,1.36]$
(transformation zone), $0.91\ [0.69,\break 1.18]$ (CIN) and $0.69\
[0.52,0.91]$ (invasive carcinoma), and $\hat{\gamma}=19.7\
[12.0,32.6]$. However, for reasons discussed above, there are drawbacks
to a conditional model, which in any case answers a different research
question to the one of primary interest, and this simple approach also
fails in itself to take account of the stochasticity of $N$.

An alternative is to model the joint distribution of $Y_{ijk}^A$ and
$N_{ij}$ directly. The correlation can be modelled using a linked
random effect approach:
%
\begin{eqnarray}
\label{common} \log\bigl(Y_{ijk}^A\bigr) & =&
\alpha^A + \beta^A_{x_{ij}} + \lambda^A
a^A_i + b_{ij} + Z_{ijk},
\nonumber
\\
N_{ij} & \sim&1 + \operatorname{Poisson}(\mu_{ij}),
\\
\mu_{ij} & =& \alpha^N + \beta^N_{x_{ij}}
+ \lambda^N a^N_i,
\nonumber
\end{eqnarray}
where $b_{ij}$ and $Z_{ijk}$ are as previously and, independently of
$b_{ij}$ and $Z_{ijk}$, $\av_i=(a^A_i,a^N_i) \sim\mathrm
{N}(0,\Sigmav
)$, where
\[
\Sigmav= \pmatrix{ 1 & \rho
\vspace*{2pt}\cr
\rho& 1}.
\]
This formulation closely follows the general three-outcome model for
one continuous, one binary and one count variable discussed by \citet
{catalanoryan1992} and \citet{ga2001}, and developed further by \citet
{gueorguieva}. The shared random effect structure considered by these
authors is very similar to~(\ref{common}), although in place of the
Poisson model for $N_{ij}$ they instead use a continuation ratio probit
model, which in our notation would take the form
\[
\mathrm{P}(N_{ij}=n|x_{ij},a_i) = \Phi\bigl(
\delta_n - \beta^N_{x_{ij}} - \lambda^N
a^N_i\bigr) \prod_{h=1}^{n-1}
\bigl\{1 - \Phi\bigl(\delta_h - \beta ^N_{x_{ij}}
- \lambda^N a^N_i\bigr)\bigr\},
\]
introducing a potentially large number of additional parameters $\delta_h$.

The work of \citet{gueorguieva} extends the model of \citet{dunson}, who
provide an integral expression for the correlation between the two
outcomes conditional on the random effects. An advantage of this joint
approach over the conditional one is that it gives an estimate of the
direct effect of tissue type on each outcome variable, similar to that
used in joint longitudinal and survival modelling [\citet{ibrahim}].

It is convenient to use a Bayesian framework to fit model (\ref
{common}). Priors were specified as follows: for each $\alpha$ and
$\beta$ parameter, $\mathrm{N}(0,10^{-6})$; for each precision
parameter, $\Gamma(10^{-3},10^{-3})$; $\lambda\sim\operatorname
{Unif}(-10,10)$; $\rho\sim\operatorname{Unif}(-0.95,0.95)$. The prior for
$\rho$ was chosen with two considerations in mind: to allow that the
outcomes might be strongly, but not perfectly, correlated, and to
ensure that draws from the distribution of $\rho$ do not allow $\Sigma$
to become singular. The model was fitted in WinBUGS v1.4 [\citet{lunn}],
with a burn-in period of 50,000 iterations. Posterior estimates were
obtained from a further 50,000 iterations, with a thinning factor of~20.
In additional analyses to check the sensitivity to the choice of
priors, the prior distributional forms and/or numerical values of the
hyperparameters were varied (although always remaining ``vague'', in the
sense of having high variance).

For comparison, this model was also fitted with $\rho$ set to zero,
equivalent to fitting univariate models to each of the outcomes in the
Bayesian framework.

\subsection{Results}

Table~\ref{jointresultstable} shows parameter estimates (median of
posterior distribution and 95\% credible interval) from model (\ref
{common}). Fitting the model with $\rho$ set to zero gives
near-identical results to Table~\ref{resultstable}, the only difference
being the LVD Transformation zone parameter, for which the estimate in
the Bayesian model was $2.37\ [1.96, 2.84]$. This allows direct
comparison of the univariate and joint results. After thinning, the
autocorrelation of posterior samples for all parameters was negligible,
and there was no material difference in the results according to the
choice of priors.

Point estimates for the $\beta$ parameters are similar to those in
Table~\ref{resultstable}, but in many cases credible intervals are
somewhat wider. The negative estimate of $\lambda^N$ was expected and
is due to the negative correlation between LVD and vessel area.
Similarly, the estimate of $\rho$ is negative, and suggests high
correlation between the individual-level random effects. The posterior
distributions of most parameters, including the components of the
random effects $b$, are approximately symmetric. An exception is $\rho
$, by virtue of being bounded through the support of the prior by $\pm
0.95$, which therefore has a positively skewed posterior distribution.

\begin{figure}

\includegraphics{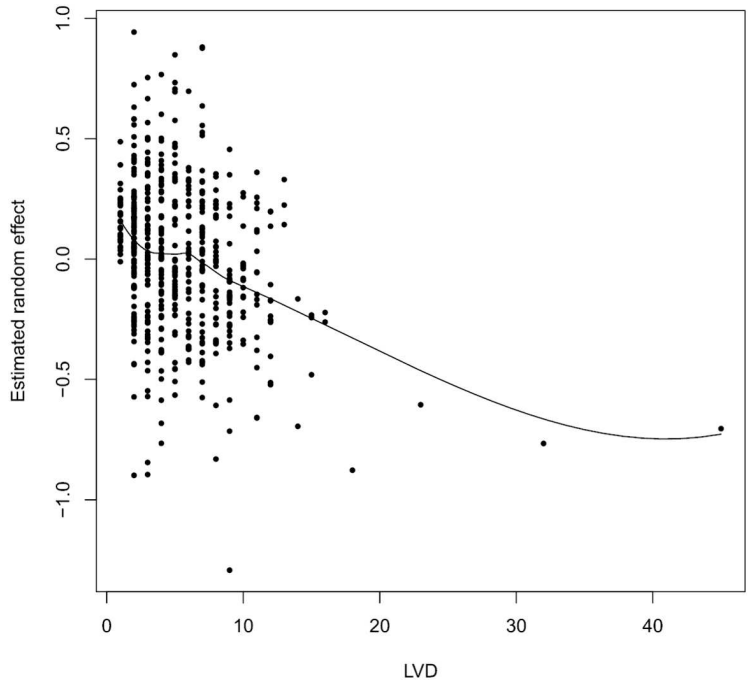}

\caption{Relationship between estimated vessel-level random effects
($b_{ij}$) from model (\protect\ref{VAeqn}) and LVD, with fitted
loess curve.}
\label{REfigure}
\end{figure}

Figure~\ref{REfigure} shows the relationship between the fitted
field-level random effects from model (\ref{VAeqn}) and LVD. This
provides a possible reason why the point estimates of the joint model
are largely unchanged compared to those from the single-model, as the
relationship is strikingly similar to that between log-vessel area and
LVD (Figure~\ref{VAversusLVD}). In the univariate model the correlation
between the cluster size and the outcome is implicitly accounted for
through the random effects distribution, even though no such
association has been specified in the formulation of the model.

\begin{table}
\caption{Parameter estimates for joint model
for LVD and vessel area}
\label{jointresultstable}
\begin{tabular*}{\textwidth}{@{\extracolsep{\fill}}lcc@{}}
\hline
& \textbf{LVD} & \textbf{Vessel area} \\
& $\bolds{\exp(\hat{\beta})}$ & $\bolds{\exp(\hat{\beta})}$ \\
\hline
Control cervix & \\
\quad Ectocervix & -- & --\\
\quad Transformation zone & 2.35 $[1.97,2.85]$ & 0.54 $[0.40,0.75]$ \\[3pt]
CIN & 2.34 $[1.96,2.78]$ & 0.42 $[0.31,0.57]$\\[3pt]
Invasive carcinoma & 3.78 $[3.17,4.47]$ & 0.26 $[0.20,0.36]$ \\[6pt]
$\hat{\lambda^A}$ & \multicolumn{2}{c}{$0.25\ [0.16,0.35]$\phantom{0.}} \\
$\hat{\lambda^N}$ & \multicolumn{2}{c}{$-0.13\ [-0.18,-0.08]$} \\
$\hat{\nu}^2$ & \multicolumn{2}{c}{$0.19\ [0.14,0.25]$\phantom{0.}} \\
$\hat{\sigma}^2$ & \multicolumn{2}{c}{$1.01\ [0.98,1.04]$\phantom{0.}} \\
$\hat{\rho}$ & \multicolumn{2}{c}{$-0.78\ [-0.92,-0.52]$}\\
\hline
\end{tabular*}
\end{table}

\section{Discussion}\label{sec5}
The analysis presented provides clear evidence of differences in
properties of lymphatic vessels according to presence of CIN or
invasive carcinoma in the cervix. Relative to the other tissue types
considered, tissue taken from individuals with invasive carcinoma have
lymphatic vessels that are greater in number, smaller in size and less
regular in shape. In addition, this study found a difference between
the ectocervix and transformation zone of the control cervix, whereas
no difference was found in LVD between control transformation zone and
CIN groups. The main scientific conclusions were relatively unaffected
by the decision to model the data in a joint as opposed to a univariate
framework.

These findings are consistent with a model that asserts that
lymphangiogenesis occurs when the cervix undergoes eversion at puberty,
to which all experimental groups would be subject, but also suggest
that there is a separate lymphangiogenenic episode in the squamous
carcinoma group. Further investigation is needed to ascertain whether
the expression of genes encoding growth factors that may cause this
additional lymphangiogenesis occurs late in CIN or after the
progression to squamous cell carcinoma. Also, the average area of
vessels in the CIN group was unexpectedly smaller than those in the
transformation zone, a finding not explained by this model of
lymphangiogenesis, confirming the need for further studies. In the
carcinoma group, vessels were considerably smaller than in all other
groups, supporting the hypothesis of newer formation, and there was
some evidence that they are morphologically and therefore functionally
different.

The approach to statistical analysis was guided by the multilevel
structure of the data set, in which it was required to account for high
within-specimen correlation in the outcome variables measured. The
negative correlation between LVD and vessel area suggested a joint
model for these two outcomes would provide a measure of the direct
effect of tissue type on the two outcomes. As one of these variables
was equal to the cluster-level sample size, a different analysis
strategy was required than that typically used, for example, for two
continuous outcome variables.

Our method used shared random effects in a bivariate Normal--Poisson
framework to allow for this correlation. We used a Bayesian framework,
although similar models have been considered utilising direct
likelihood maximisation [\citet{gueorguieva}]. Allowing for the
randomness in cluster size tended to increase the standard errors of
parameter estimates, but there were no substantial changes in the point
estimates themselves.

This finding agrees with the results of \citet{neuhaus}, who demonstrate
that for linear mixed models with random intercept only, estimators of
covariate effects are consistent, and are estimated equally
efficiently, even when cluster size is ignored. They point out that the
exception is the fixed intercept parameter ($\alpha$), which in any
case is rarely of substantive interest. However, they assume that
cluster sizes do not depend on covariates, which is not plausible in
our application in that cluster size itself is an outcome measure, with
strong evidence of an association with tissue type.

Additionally, the data considered here did not warrant using random
slope models, which have been used successfully elsewhere when
continuous covariates were available [e.g., \citet{dunson}]. In general,
the area of informative cluster size has received little attention in
the statistics literature, and has only been applied to a limited
number of application areas, which is perhaps surprising given the
rapidly increasing level of research in the area of cluster randomised trials.

The joint modelling framework considered could in principal be extended
to the multivariate case. In our analysis this was not necessary, as
there was no reason why circularity should be associated with the
number or size of vessels. To achieve this, similar models might be
considered that use several shared random effect terms to induce
correlation between the random variables. \citet{catalanoryan1992}
provide further details.

This study did not attempt to distinguish between the cancer grade or
stage of samples analysed, and also makes no link between properties of
lymphatic vessels and metastasis or prognosis. These are both
limitations and possible future research directions, and it would be a
simple extension of the joint framework set out here to analyse data of
this type. In addition, in this study it was not possible to measure
either the spatial distribution of lymphatic vessels within a specimen
or changes in lymph structure relative to distance from the tumour
site, both of which may provide extra insight.

\section*{Acknowledgments}
This manuscript has benefitted from the comments of a reviewer, an
Editor and an Associate Editor.



%


\printaddresses
\end{document}